# Leidenfrost temperature: surface thermal diffusivity and effusivity effect

Omar Lamini, Rui Wu*, C.Y. Zhao**

Institute of Engineering Thermophysics, Shanghai Jiao Tong University, 800 Dongchuan Road, Shanghai 200240, China

**ABSTRACT**

Droplet impact on hot surfaces results in either droplet-surface contact or droplet-surface non-contact, i.e., the Leidenfrost state. The Leidenfrost droplet is levitated upon its vapor, deteriorating the heat transfer. The Leidenfrost temperature depends on the thermal properties of the surface, which are usually characterized by two parameters: the thermal diffusivity and the thermal effusivity. In this paper, the effects of these two parameters on the Leidenfrost temperature are clarified experimentally by varying the one of interest while keeping the other one constant. The experimental results indicate that the Leidenfrost temperature is affected by the thermal effusivity rather than by the thermal diffusivity; the higher the thermal effusivity is, the lower the Leidenfrost temperature; and the increase of the Leidenfrost temperature with the droplet impact velocity is greater for the surface with the lower thermal effusivity. To further understand the experimental findings, a theoretical model is proposed, which considers the transient heat transfer in the surface. The theoretical analysis shows that the Leidenfrost temperature scales as the inverse of the thermal effusivity and the root square of the impact velocity, in good agreement with the experimental results.

**INTRODUCTION**

When a liquid droplet impacts on a surface above a critical temperature, i.e., the so called Leidenfrost temperature, the droplet levitates upon its own vapor, known as the Leidenfrost phenomenon [1]. Tuning the Leidenfrost temperature is of vital interest for a wide range of applications such as spray cooling [2-3], fuel injection [4], and drag reduction [5-6]. The Leidenfrost temperature depends on the structure and thermal properties of the surface, but the dependency is still unclear [7]. Both the increase [8-11] and the decrease [12-13] of the Leidenfrost temperature have been reported for the engineered surfaces with porous coatings, micropillars, and nanowires. In fact, the structures of solid surfaces influence the thermal properties of the solid as well [14]. Hence, in order to tune the Leidenfrost temperature, it is necessary to understand in detail the effects of surface thermal properties.

Both the thermal diffusivity and effusivity have been proposed to account for the effects of the surface thermal properties on the Leidenfrost temperature. The surface thermal diffusivity is $\alpha = k/\rho C_p$ and the thermal effusivity is $e = \sqrt{k\rho C_p}$, where $k$ is the surface thermal conductivity, $\rho$ is the material density and $C_p$ is the specific heat capacity. In the study of Baumeister and Simon [15], a parameter inverse of the square of the thermal effusivity was used. Orejon et al. [16] deemed that it is more relevant to use the thermal diffusivity and showed that the Leidenfrost temperature decreases by increasing the surface thermal diffusivity. Nevertheless, Emmerson [17] revealed that the Leidenfrost temperature is not dependent on the thermal diffusivity. As can be seen, different conclusions were reached for the role of the thermal diffusivity and effusivity.

One possible reason for this contradiction could be that in the above-mentioned studies, the thermal diffusivity and effusivity were not controlled independently, and the influence of these two parameters on the Leidenfrost temperature cannot be surmised.

To distinguish the influence of the thermal diffusivity and effusivity in the present study, we investigate experimentally the effects of these two parameters on the Leidenfrost temperature by varying the one of interest while keeping the other one constant. We find that the Leidenfrost temperature depends on the thermal effusivity rather than the thermal diffusivity. A theoretical model is also developed, which takes into account the transient heat transfer in the surface. The theoretical analysis shows that the Leidenfrost temperature scales as the inverse of the thermal effusivity and the root square of the impact velocity, in good agreement with the experimental results.

**EXPERIMENT**

To disclose the effects of the thermal diffusivity, we use the surfaces with similar thermal effusivity but different thermal diffusivity: Aluminum (99.9% Al), Brass (85% Cu and 15% Zn), and Armco iron (99.8% Fe). On the other hand, Armco iron (99.8% Fe) and Antimonial lead (94% Pb and 6% Sb), which have similar thermal diffusivity but different thermal effusivity, are employed to study the influence of the thermal effusivity. The properties of these materials are shown in Table 1. All the surfaces are 40 mm diameter and 15 mm thick and gradually polished with sandpaper (the AFM measurement shows that the surface roughness is less than 0.05 μm). Six heater cartridges (48 W each) and a K-type thermocouple inserted 1 mm beneath the surface (Keysight 34970A LXI) are used to control the surface temperature, see Figure 1a.

The experimental setup for the droplet impact on hot surfaces consisted of a heating system, a droplet generator, a high-speed camera, and data acquisition system, see Figure 1b. Droplets with diameter of $D$=2.7 mm are generated by injecting ethanol into a blunt needle with an inner diameter of 0.86 mm. The injection rate of ethanol is 0.1 ml/min and controlled by a syringe pump (11 Elite Harvard apparatus). The needle is heightened from 3.5 to 65 cm, resulting in the impact velocity $U$ from 0.73 to 3.28 m/s and the Weber number from 51 to 1030. The Weber number is defined as $We = \rho_l U^2 D/\sigma$, where $\rho_l$ is the droplet liquid density, $U$ is the droplet impact velocity and $\sigma$ is the droplet liquid surface tension. A high-speed camera (IDT NXA7-S1) is used to record the droplet impact processes, with a temporal resolution of 2000 fps and a spatial resolution of 15.3 μm/pixel.

**RESULTS AND DISCUSSION**

Five distinct droplet impact regimes are observed in the experiments: deposition with secondary atomization, rebound with secondary atomization, breakup with secondary atomization, rebound, and breakup, as shown in Figure 2. The detailed description of these five regimes can be found in Bertola [18]. The Leidenfrost temperature is defined as the minimum temperature to observe the rebound or breakup without the secondary atomization.

The regimes of droplet impact on different surfaces are shown in Figure 3. The Leidenfrost temperature is about 150 ℃ for aluminum and brass, and 160 ℃ for Armco iron (these three surfaces have different thermal diffusivity but similar thermal effusivity), Figure 3(a-c). It indicates that the thermal diffusivity has little effect on the Leidenfrost temperature. This result, however, contradicts the study of Orejon et al. [16], in which it was suggested that the Leidenfrost temperature is influenced by the thermal diffusivity. Nevertheless, in the study of Orejon et al. [16], the surfaces have different thermal diffusivity and effusivity, and hence we cannot distinguish which thermal property influences the Leidenfrost temperature.

Since the melting temperature of antimonial lead is not more than 190 ℃, the maximum temperature tested on this surface is 170 ℃. At such maximum temperature, most of the droplet impacts lie in the deposition with secondary atomization regime, Figure 3(d), indicating that the Leidenfrost temperature is larger than 170 ℃. By contrast, the Leidenfrost temperature of Armco iron is about 160 ℃, which has a similar thermal diffusivity but different thermal effusivity as antimonial lead. In the experiment, the increment of surface temperature is 5 ℃ when the Leidenfrost state is approaching. Hence, the error of the Leidenfrost temperature measured is not more than 5 ℃, which indicates that the Leidenfrost temperature of antimonial lead is larger than that of Armco. However, it is not easy to get more surfaces with similar thermal diffusivity but different thermal effusivity as Armco iron and/or antimonial lead. That is why only these two surfaces are used to investigate the effect of thermal diffusivity.

To further understand the effects of the thermal effusivity on the Leidenfrost temperature, we compare in Figure 4 the Leidenfrost temperature of the following liquid-surface combinations obtained from the present and previous experimental studies: ethanol-aluminum, ethanol-sapphire [19] and ethanol-glass [20]; water-aluminum [18], water-silicon [13], water-sapphire [21] and water-zirconium [22]. Although these surfaces have different thermal diffusivities and effusivities, it is reasonable to employ them to reveal the effects of the thermal effusivity, since the role of the thermal diffusivity, as discussed above, is not significant. As shown in Figure 4, the Leidenfrost temperature decreases with the increase of the surface thermal effusivity.

Based on the present experimental results (Figure 3) and the previous studies (Figure 4), it can be inferred that the Leidenfrost temperature depends on thermal effusivity rather than the thermal diffusivity.

The results in Figure 3 also show that the increase in the Leidenfrost temperature with the droplet impact velocity is small, not more than 10 °C for all surfaces. The previous studies [13, 18-20, 23, 24] have shown various increment in the Leidenfrost temperature with the increase of the droplet impact velocity for different droplets and surfaces. In Figure 5, we show the variation of the Leidenfrost temperature, $T_{L,v}$, as the impact velocity increases from 0.73 and 3.28 m/s for ethanol droplet with various surfaces. For the surface with lower thermal effusivity, $T_{L,v}$ is larger.

The thermal effusivity elucidates how fast a material is able to transfer heat from the surface to the environment [25]. Hence, for the surfaces with higher thermal effusivity, more heat will be transferred to the droplet, leading to more generated vapor from the bottom of the droplet. As a result, the pressure in the gas film between the droplet and the surface becomes higher, rendering a lower Leidenfrost temperature, and a smaller increase of the Leidenfrost temperature with the droplet impact velocity.

To understand in detail the effects of thermal effusivity on the Leidenfrost temperature, the following theoretical model is proposed, which considers the transient heat transfer in the surface. The schematic of a Leidenfrost droplet is shown in Figure 6. The heat transfer in the surface is considered as the one-dimensional transient heat conduction in a semi-infinite solid:

$$\frac{\partial^2 T}{\partial x^2} = \frac{1}{\alpha}\frac{\partial T}{\partial t} \qquad (1)$$

At the solid surface, $x = 0$, and at the solid bottom, $x = \infty$. Initially, the solid temperature is $T_i$.

The temperature at the bottom of the droplet is assumed to be the saturation temperature $T_{sat}$ [26-27]. The heat flux from the solid surface to the droplet is expressed as:

$$q = \frac{e_s}{\sqrt{\pi t}}(T_i - T_s) = \frac{k_{film}}{h_{film}}(T_s - T_{sat}) \qquad (2)$$

where $e_s$ is the thermal effusivity of solid, $T_s$ the temperature of the solid surface, $k_{film}$ and $h_{film} = h_{film}(r,t)$ are the thermal conductivity and the thickness of the gas film between the droplet and the surface, respectively. From Eq. (2), we can get the heat flux:

$$q = \frac{T_i - T_{sat}}{\frac{\sqrt{\pi t}}{e_s} + \frac{h_{film}}{k_{film}}} \qquad (3)$$

Based on the work of Klaseboer et al. [28], the pressure of the gas film is:

$$P_{film} - P_0 = \frac{2\sigma}{R_T} + \frac{2\sigma}{R_B} + \frac{1}{2}\rho_l U^2 + \rho_l g h_d \qquad (4)$$

where $P_0$ is the environmental pressure and is constant, $\sigma$ the surface tension, $U$ the droplet impact velocity, $h_d$ the droplet height, $R_T$ and $R_B$ the radii of curvatures at the top and bottom of the droplet, respectively.

The time evolution of the thickness of the gas film between the droplet and the surface is expressed as:

$$\frac{\partial h_{film}}{\partial t} = \frac{1}{12\mu_{film}}\frac{1}{r}\frac{\partial}{\partial r}\left(r h_{film}^3 \frac{\partial P_{film}}{\partial r}\right) + \frac{q}{L\rho_v} \qquad (5)$$

where $\mu_{film}$ is the dynamic viscosity of the gas in the film, $L$ the latent heat of vaporization, $\rho_v$ the vapor density. It should be noted that Eq. (5) is valid only if the Reynolds number for the flow in the gas film is.

$$Re = \frac{\rho_{film} U (RH)^{1/2}}{\mu_{film}} \ll 1 \qquad (6)$$

where $\rho_{film}$ is the density of gas in the film, $H$ the characteristic film thickness, and radial dimension $r \sim (RH)^{1/2}$. $R$ is the radius of the spherical droplet.

Based on Eq. (4), the characteristic pressure of the gas film is defined as:

$$P = \frac{\sigma}{R} + \rho_l U^2 + \rho_l gR \tag{7}$$

For the Leidenfrost droplet, the time evolution of the film thickness can be rather small, i.e., $\partial h_{film}/\partial t \approx 0$. Hence, the values of the two terms at the right hand of Eq. (5) are in the same order:

$$\frac{1}{\mu_{film}} \frac{H^3 P}{HR} \sim \frac{q}{L\rho_v} = \frac{T_i - T_{sat}}{\frac{\sqrt{\pi t}}{e_s} + \frac{H}{k_{film}}} \frac{1}{L\rho_v} \tag{8}$$

By using Eqs. (3) and (7) as well as $t \sim R/U$, we can get:

$$T_i - T_{sat} \sim \left(\frac{\sqrt{\pi R}}{e_s\sqrt{U}} + \frac{H}{k_{film}}\right) \frac{L\rho_V}{\mu_{film}} \frac{H^3}{HR} \rho_l U^2 \left(\frac{1}{We} + 1 + \frac{Bo}{We}\right) \tag{9}$$

where $We = \rho_l R U^2/\sigma$ is the Weber number, and $Bo = \rho_l g R^2/\sigma$ is the Bond number.

From Eq. (9) we can obtain the influence of the surface thermal effusivity $e_s$: $T_i \sim 1/e_s$. To this end, it can be concluded that the Leidenfrost temperature follows:

$$T_L \sim \frac{1}{e_s} \tag{10}$$

The Leidenfrost temperature predicted by Eq. (10) is also shown in Figure 4 (see the line). The predicted results agree well with the experimental data.

For the high impact velocity with $We \gg 1$, it has been revealed that [28]:

$$\frac{H}{R} \sim \sqrt{\frac{\mu_{film}}{\rho_l U R}} \tag{11}$$

Substituting Eq. (11) into Eq. (9) and neglecting $1/We$ and $Bo/We$ due to the high Weber number, we can get:

$$T_i - T_{sat} \sim \left(\frac{\sqrt{\pi R}}{e_s} + \frac{\sqrt{R\mu_{film}}}{k_{film}\sqrt{\rho_l}}\right) L\rho_V \sqrt{U} \qquad (12)$$

Form Eq. (12), we can get:

$$T_L \sim U^{1/2} \qquad (13)$$

The results predicted by Eq. (13) are also shown in Figure 3 (see the line). The predicted and experimental results are in good agreement for all tested surfaces.

The agreement between the experimental data and the results predicted by Eqs. (10) and (13) demonstrates the effectiveness of the present theoretical model. In addition, Eq. (9) indicates that the increase of the Leidenfrost temperature with the increase of the droplet impact velocity is inverse proportional to the surface thermal effusivity, also in agreement with the experimental results shown in Figure 5.

**CONCLUSION**

In summary, the effect of the surface thermal diffusivity and effusivity on the Leidenfrost temperature is clarified. It is revealed experimentally that the Leidenfrost temperature is affected by the thermal effusivity rather than by the thermal diffusivity; the higher the thermal effusivity is, the lower the Leidenfrost temperature; and the increase of the Leidenfrost temperature with the droplet impact velocity is greater for the surface with the lower thermal effusivity. A theoretical model, which considers the transient heat transfer in the surface, is developed to shed light on the role of the thermal effusivity. The model indicates that the Leidenfrost temperature scales as the inverse of the surface thermal effusivity and the root square of the droplet impact velocity, in good agreement with the experimental results. This agreement not only demonstrates the effectiveness of the developed model in the present study but also implies that the transient heat transfer in the surface needs to be considered so as to gain accurate description of the Leidenfrost temperature.

ACKNOWLEDGMENT

This work is supported by the National Key Research and Development Program of China (No. 2018YFC1800600), the National Natural Science Foundation of China (No. 51776122), the Shanghai Pujiang Program (No. 17PJ1404600), and the Shanghai Key Fundamental Research Grant (Grant Nos. 18JC1413300, 16JC1403200).

21. van Limbeek, M. A. J.; Hoefnagels, P. B. J.; Sun, C.; Lohse, D., Origin of spray formation during impact on heated surfaces. *Soft Matter* **2017,** *13* (41), 7514-7520.
22. Lee, G. C.; Kang, J.-y.; Park, H. S.; Moriyama, K.; Kim, S. H.; Kim, M. H., Induced liquid-solid contact via micro/nano multiscale texture on a surface and its effect on the Leidenfrost temperature. *Experimental Thermal and Fluid Science* **2017,** *84*, 156-164.
23. Staat, H. J. J.; Tran, T.; Geerdink, B.; Riboux, G.; Sun, C.; Gordillo, J. M.; Lohse, D., Phase diagram for droplet impact on superheated surfaces. *Journal of Fluid Mechanics* **2015,** *779*.
24. Tran, T.; Staat, H. J.; Prosperetti, A.; Sun, C.; Lohse, D., Drop impact on superheated surfaces. *Physical review letters* **2012,** *108* (3), 036101.
25. Marń, E., Thermal Physics Concepts: The Role of the Thermal Effusivity. *The Physics Teacher* **2006,** *44* (7), 432-434.
26. van Limbeek, M. A. J.; Klein Schaarsberg, M. H.; Sobac, B.; Rednikov, A.; Sun, C.; Colinet, P.; Lohse, D., Leidenfrost drops cooling surfaces: theory and interferometric measurement. *Journal of Fluid Mechanics* **2017,** *827*, 614-639.
27. Sobac, B.; Rednikov, A.; Dorbolo, S.; Colinet, P., Leidenfrost effect: Accurate drop shape modeling and refined scaling laws. *Phys Rev E Stat Nonlin Soft Matter Phys* **2014,** *90* (5-1), 053011.
28. Klaseboer, E.; Manica, R.; Chan, D. Y., Universal behavior of the initial stage of drop impact. *Physical review letters* **2014,** *113* (19), 194501.
**AUTHOR INFORMATION**

**Corresponding author:**

* E-mail: ruiwu@sjtu.edu.cn

** E-mail: changying.zhao@sjtu.edu.cn
**Notes:**

The authors declare no competing financial interest.

# LIST OF TABLES

**Table 1:** Thermophysical properties of the surfaces used in the present experiments.

**LIST OF FIGURES**

**Figure 1:** Schematic of (a) heater surface and (b) experimental setup

**Figure 2:** Various ethanol droplet impact dynamics on the hot Armco iron surface; (a) deposition with secondary atomization at the surface temperature of 140 ℃ and the impact velocity of 0.73 m/s;  (b) rebound with secondary atomization at the surface temperature of 150 ℃ and the impact velocity of 0.73 m/s; (c) breakup with secondary atomization at the surface temperature of 145 ℃ and the impact velocity of 2.48 m/s; (d) rebound at the surface temperature of 165 ℃ and the impact velocity of 0.73 m/s; (e) breakup at the surface temperature of 165 ℃ and the impact velocity of 2.48 m/s.

**Figure 3:** Regime maps for the ethanol droplet impact on (a) aluminum, (b) brass, (c) Armco iron and (d) antimonial lead

**Figure 4:** The Leidenfrost temperature versus the surface thermal effusivity

**Figure 5:** Variation of the Leidenfrost temperature versus the surface thermal effusivity. The bars represent the variation of the Leidenfrost temperature for ethanol droplet on various surfaces between the impact velocities of 0.73 and 3.28 m/s.

**Figure 6:** Schematic of a Leidenfrost droplet

**Table 1**

| Surface material | Thermal conductivity $k$ (W/m K) | Specific heat capacity $C_p$ (J/kg K) | Material Density $\rho$ (kg/m³) | Thermal diffusivity $\alpha$ (mm²/s) | Thermal effusivity $e$ (kW s$^{0.5}$/m² K) |
|---|---|---|---|---|---|
| Aluminum | 226 | 921 | 2698 | 91 | 23.688 |
| Brass | 146 | 377 | 8750 | 44 | 21.967 |
| Armco iron | 74 | 850 | 7860 | 20 | 22.200 |
| Antimonial lead | 30 | 134 | 10960 | 20 | 6.603 |

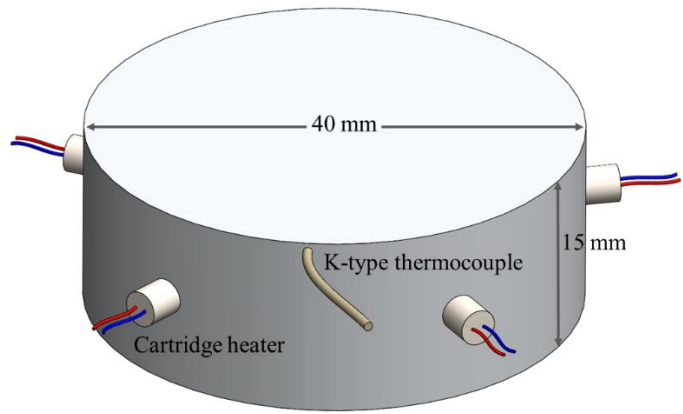

(a)

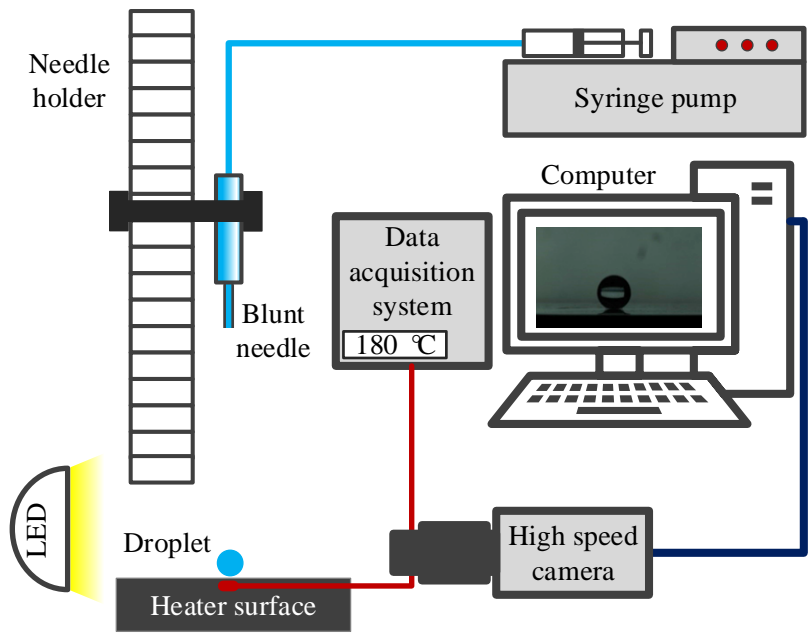

(b)

**Figure 1**

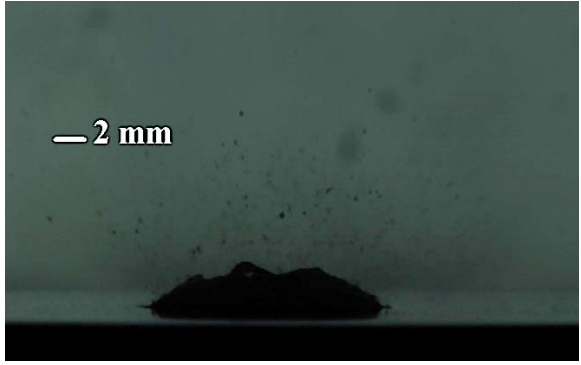 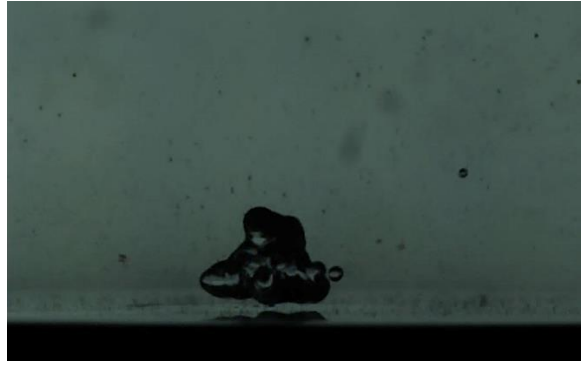

(a)　　　　　　　　　　　　　　　　　　　(b)

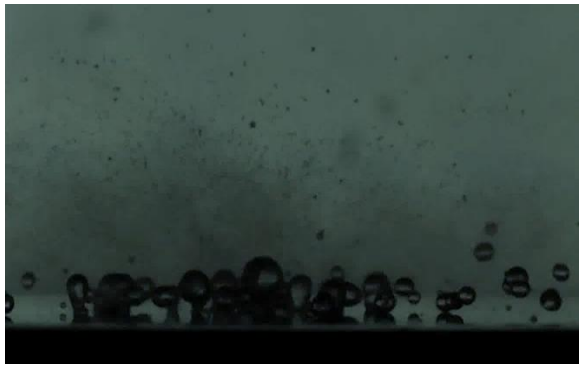 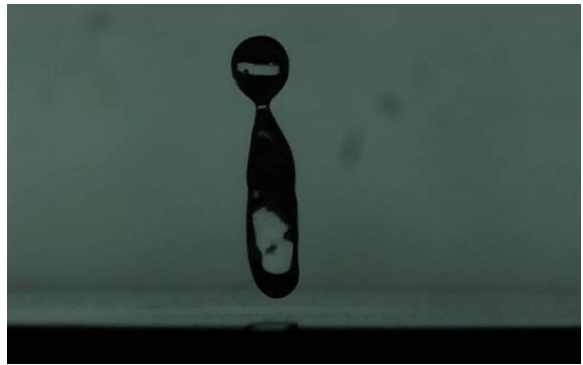

(c)　　　　　　　　　　　　　　　　　　　(d)

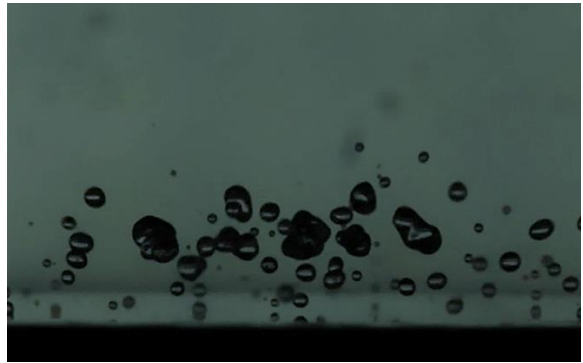

(e)

**Figure 2**

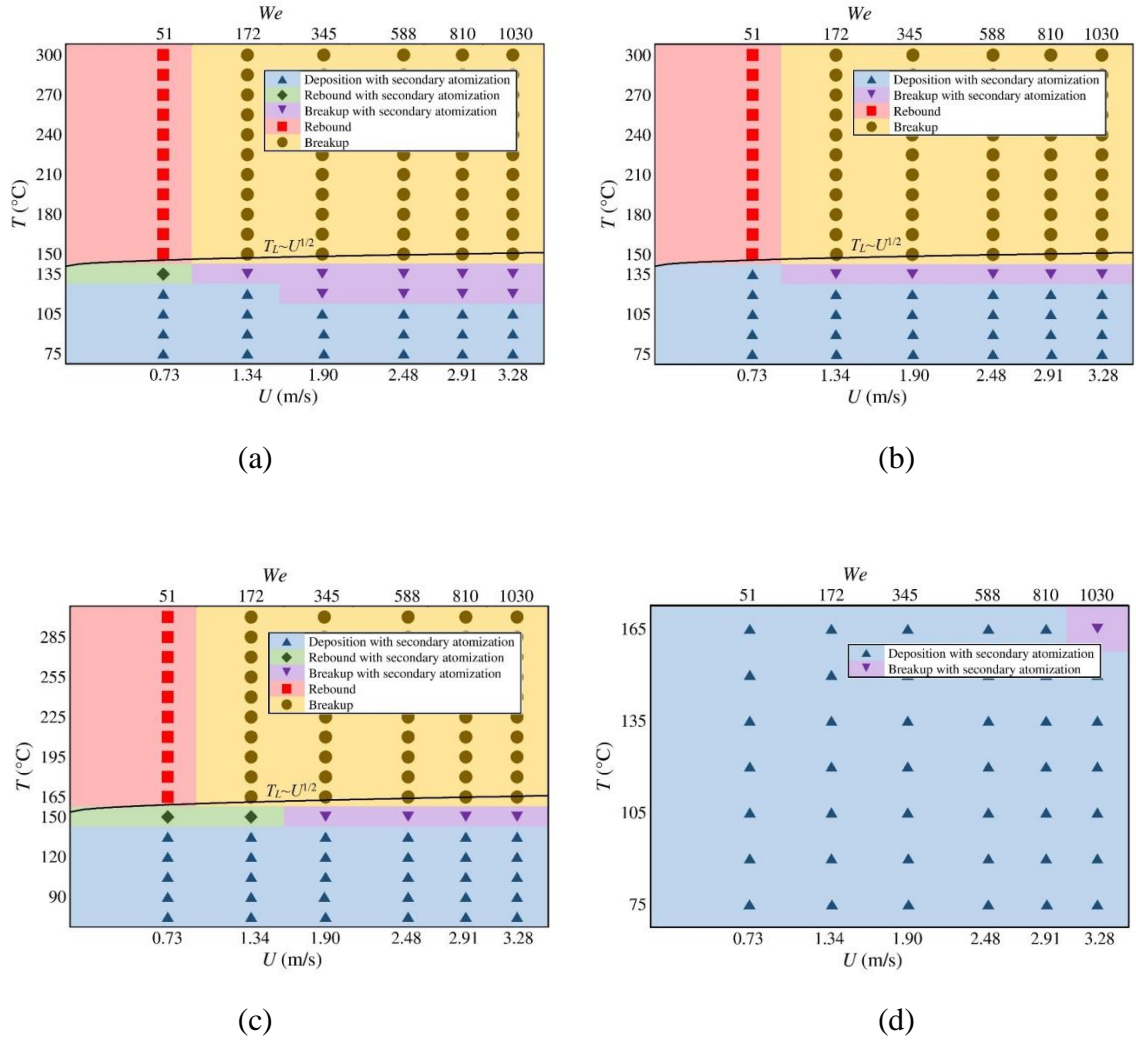

**Figure 3**

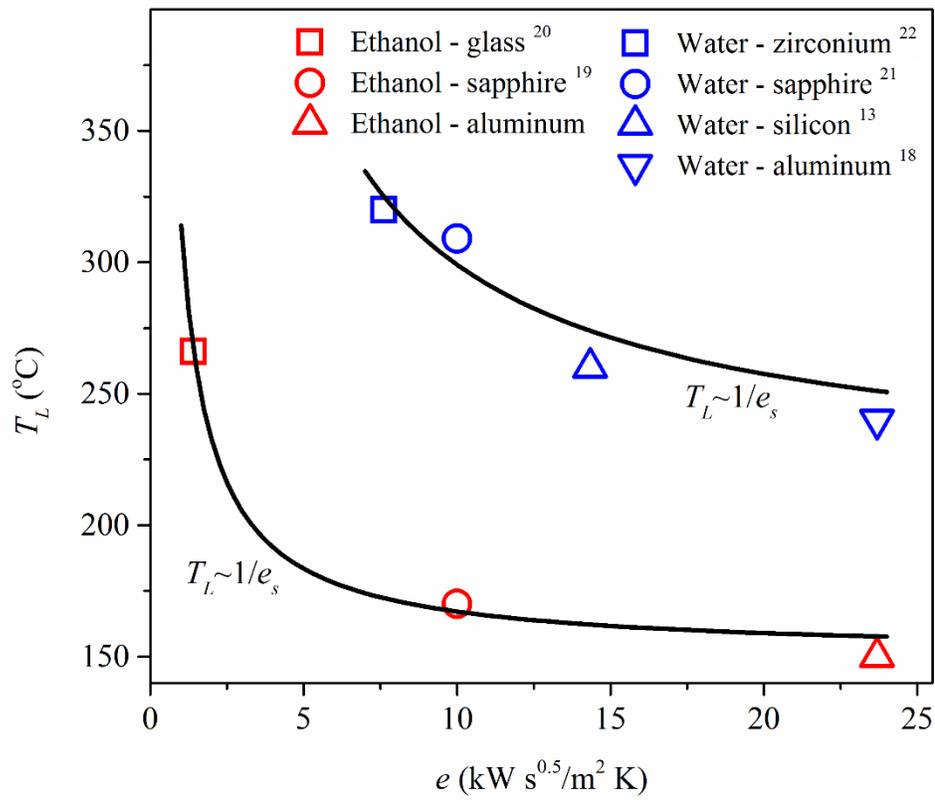

**Figure 4**

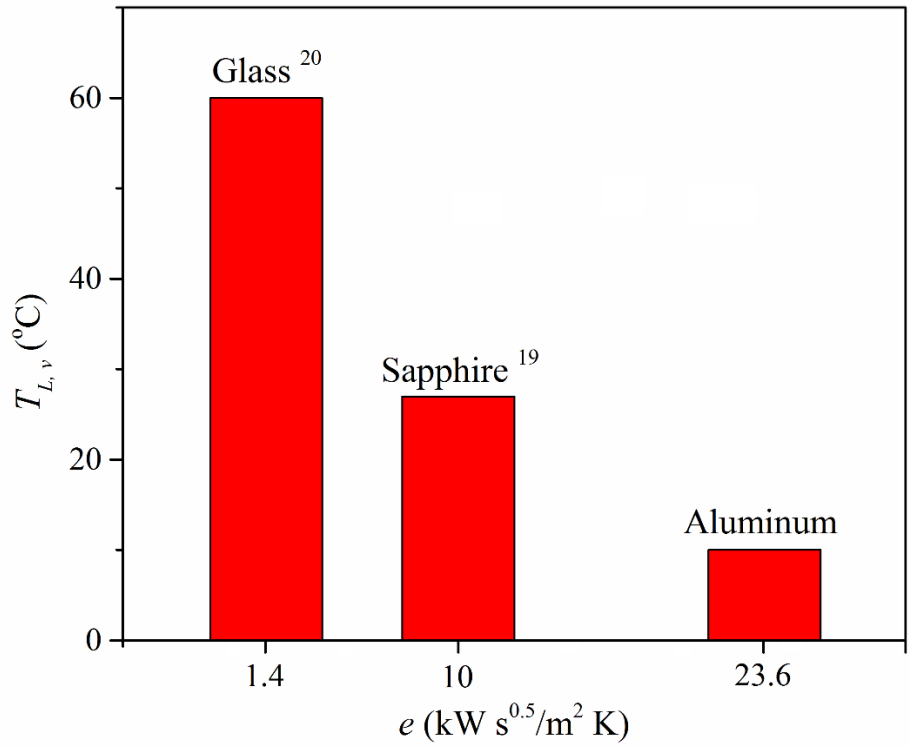

**Figure 5**

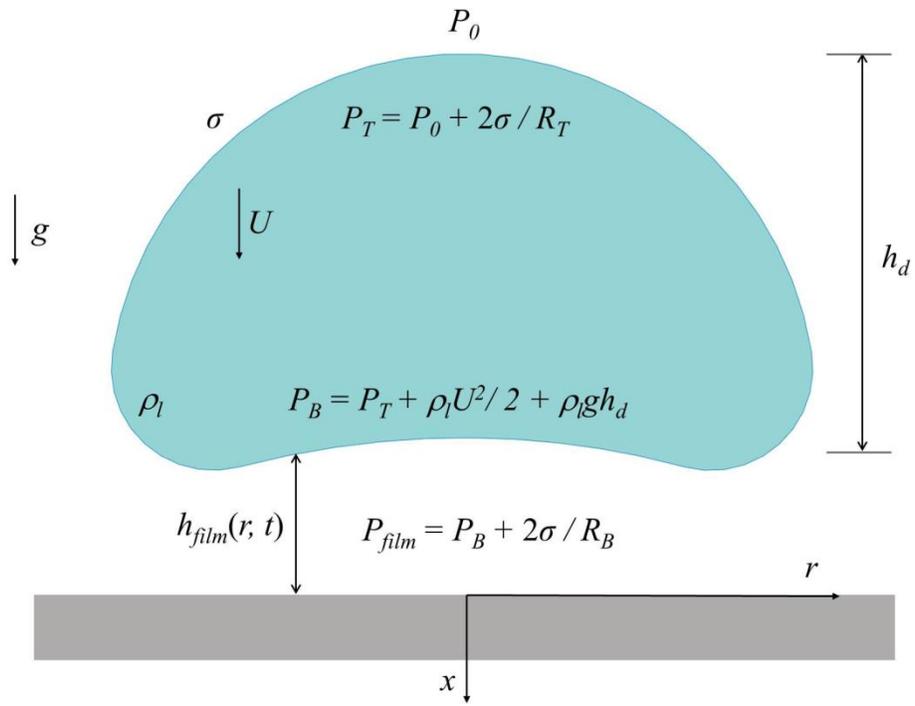

**Figure 6.**